\documentclass[trackchanges,twocolumn]{aastex}

\usepackage{natbib,epsfig,siunitx,url,graphicx,amsmath,footnote,float,xcolor,cases}
\begin{document}

\submitted{Icarus, accepted}

\title{Rethinking the role of the giant planet instability in terrestrial planet formation models}

\author{Matthew S. Clement\altaffilmark{1,2}, Rogerio Deienno\altaffilmark{3} \& Andr\'{e} Izidoro\altaffilmark{4}}

\altaffiltext{1}{Earth and Planets Laboratory, Carnegie Institution for Science, 5241 Broad Branch Road, NW, Washington, DC 20015, USA}
\altaffiltext{2}{Johns Hopkins APL, 11100 Johns Hopkins Rd, Laurel, MD 20723, USA}
\altaffiltext{3}{Southwest Research Institute, 1050 Walnut St. Suite 300, Boulder, CO 80302, USA}
\altaffiltext{4}{Department of Earth, Environmental and Planetary Sciences, MS 126, Rice University, Houston, TX 77005, USA}
\altaffiltext{*}{corresponding author email: matt.clement@jhuapl.edu}

\begin{abstract}

Advances in computing power and numerical methodologies over the past several decades sparked a prolific output of dynamical investigations of the late stages of terrestrial planet formation.  Among other peculiar inner solar system qualities, the ability of simulations to reproduce the small mass of Mars within the planets' geochemically inferred accretion timescale of $\lesssim$10 Myr after the appearance of calcium aluminum-rich inclusions (CAIs) is arguably considered the gold standard for judging evolutionary hypotheses.  At present, a number of independent models are capable of consistently generating Mars-like planets and simultaneously satisfying various important observational and geochemical constraints.  However, all models must still account for the effects of the epoch of giant planet migration and orbital instability; an event which dynamical and cosmochemical constraints indicate occurred within the first 100 Myr after nebular gas dispersal.  If the instability occurred in the first few Myr of this window, the disturbance might have affected the bulk of Mars' growth.  In this manuscript, we turn our attention to a scenario where the instability took place after $t\simeq$ 50 Myr.  Specifically, we simulate the instability's effects on three nearly-assembled terrestrial systems that were generated via previous embryo accretion models and contain three large proto-planets (i.e. Earth, Venus and Theia) with 0.5 $<m<$ 1.0 $M_{\oplus}$ and orbits interior to a collection of $\sim$Mars-mass embryos ($a>$1.3 au and $m<$ 0.2 $M_{\oplus}$) and debris. While the instability consistently triggers a Moon-forming impact and efficiently removes excessive material from the Mars-region in our models, we find that our final systems are too dynamically excited and devoid of Mars and Mercury analogs.  Thus, we conclude that, while possible, our scenario is far more improbable than one where the instability either occurred earlier, or at a time where Earth and Venus' orbits were far less dynamically excited than considered here.

\end{abstract}

\section{Introduction}
\label{sect:1}

As the process of forming an Earth-mass planet from $\sim \micron$-scale dust in a proto-planetary nebula involves a change of nearly 30 orders of magnitude in mass in just a few tens of millions years \citep{haisch01,hernandez07}, numerical modeling of the process is typically divided into several key phases.  These include: (1) the direct and rapid conversion of dust particles \citep{brice01} into $\sim$10-1,000 km planetesimals \citep{youdin05,johansen07,draz16,simon16}, (2) the formation of the terrestrial planets' seed embryos via runaway growth of planetesimals \citep{koko_ida_96,morishima08} and (3) the epoch of giant impacts (defined in this paper as an impactor with $m>$ 0.01 $M_{\oplus}$) that largely terminated the accretion process \citep{chambers98} and birthed the Moon \citep{asphaug14_moon}.  While this paper primarily focuses on the study of the giant impact phase (3), investigations of each individual period of planet growth broadens our understanding of the other epochs.  Thus, contemporary models increasingly strive to holistically model the complete assembly of the terrestrial worlds from the ground up \citep[e.g.:][]{woo21_tp,broz21,johansen21,izidoro21_nat}.

In classic investigations of the giant impact phase \citep{wetherill80_rev,chambers98,agnor99}, tens to hundreds of Moon-Mars-mass embryos \citep[0.01-0.1 $M_{\oplus}$; the product of runaway growth:][]{koko_ida_96} engulfed in a planetesimal sea \citep[e.g.:][]{morby09_ast,delbo17} collisionally grow over $\sim$100 Myr timescales.  While these archetypal studies proved remarkably successful at broadly replicating many qualities of the inner solar system \citep[e.g.:][]{chambers01,ray07,raymond13,fischer14}, they also demonstrated how the diminutive masses of Mercury, Mars and the asteroid belt are inconsistent with a scenario where the inner solar system originates from a uniform distribution of solids between the Sun and Jupiter \citep{agnor99,chambers_cassen02,hansen09}.  Though other discrepancies between these model generated systems and the present-day terrestrial planets \citep[among others, the low orbital eccentricities and inclinations of the planets:][]{ray06,obrien06,clement18_frag,deienno19} have received moderate focus in the recent literature, the so-called small-Mars problem \citep{ray09a} has unquestionably dominated the attention of the preceding two decades worth of investigations.

One potential explanation for Mars' low mass involves supposing that few planetesimals ever formed in the vicinity of the planets' current orbit and the asteroid belt \citep{izidoro15,ray17,ray17sci,izidoro21_tp}.  Indeed, modern models of planetesimal formation elucidate the process' sensitivity to the in-situ properties of the nebular disk \citep{simon16,draz16,johansen21,izidoro21_nat,morby22}.  If this were not the case, and planetesimal formation was indeed efficient throughout the inner solar system, excessive mass in the outer terrestrial disk could have been removed during the gas disk phase if Jupiter and Saturn migrated in and out of the region \citep[the ``Grand Tack'' model:][]{walsh11,pierens11,jacobson14} or via sweeping secular resonances \citep[the ``Dynamical Shake-up'' described in:][]{nagasawa00,thommes08,bromley17} with the primordially eccentric \citep{pierens14,clement21_instb} giant planets during disk photo-evaporation.  However, the strong radial mixing in either depletion scenario is potentially inconsistent with the modern, distinct isotopic compositions of Earth and Mars \citep{tang14,dauphas17,woo21_tp}.  This might be avoided if the inner disk was instead reshaped via convergent migration of the terrestrial material itself \citep[e.g.:][]{cresswell08,paardekooper11,bitch15,eklund17,broz21}.  Finally, if the terrestrial disk was not truncated prior to gas dispersal, Jupiter and Saturn's acquisition of their modern orbital configuration could have perturbed Mars' formation \citep{ray09a}.  \citet{clement18} studied the effects of an unusually early giant planet instability, occurring some 1-10 Myr after gas dispersal, on the forming terrestrial planets and found that Mars' mass and formation timescale are best matched in simulations that most closely reproduce Jupiter and Saturn's modern dynamical configuration \citep{clement18_ab,clement21_tp}.  However, certain geochemical constraints \citep[e.g.:][]{marty17} have been interpreted to preclude an instability occurring prior to core-closure on Earth.

With the small-Mars problem largely reduced to a debate over the efficacy of different formation models, we turn our attention to the young inner solar systems' evolution around the inferred time of the Moon-forming impact \citep[30 $\lesssim t \lesssim$ 150 Myr as inferred from isotopic dating:][]{wood05,kleine09,rudge10,kleine17,barboni17}.  While cosmochemical studies typically record time with respect to CAI formation, for the remainder of this manuscript we refer to nebular dispersal ($t_{CAI}+$ $\sim$2-4 Myr) as time zero, which we loosely correlate with the beginning of most N-body simulations of the giant impact phase.  In particular, a renewed focus on this era is increasingly relevant given a recent shift in our understanding of the solar system's giant planet instability \citep{Tsi05,levison08,nesvorny11}.  The so-called Nice Model describes how the giant planets attain their modern orbits through a violent epoch of dynamical instability characterized by strong mutual encounters and rapid evolution in the planets' semi-major axes and eccentricities.  While such a violent episode is highly likely to destabilize or over-excite the fully formed terrestrial planets \citep{bras09,bras13,kaibcham16}, it is now widely accepted that the event transpired in conjunction with the late stages of their formation. Specifically, geophysical studies of impact and differentiation chronologies in the inner solar system \citep[e.g.:][]{evans18,morb18,mojzsis19,brasser20}, geochemical measurements of meteorites and lunar samples \citep[e.g.:][]{boehnke16,zellner17,goodrich21,worsham21} and dynamical studies of the stability of fragile dynamical structures during the violent instability \citep[e.g.:][]{delbo17,nesvorny18,quarles19,ribeiro20,nesvorny21_e_nep} all support an instability that occurs within the first 100 Myr after gas disk dispersal.

The main goals of this paper are to investigate whether a Nice Model instability occurring around the time of the Moon-forming impact (at $t\lesssim$100 Myr to satisfy constraints on the instability's timing, and also within the earlier part of the $\sim$30-150 Myr window of inferred timings of the Moon-forming impact) is consistent with the terrestrial system's modern low degree of dynamical excitation \citep[e.g.:][]{roig16}, and also capable of consolidating unnecessary proto-planets in the vicinity of Earth's orbit \citep[i.e.: directly triggering the Moon-forming impact as proposed in][]{desouza21}, removing excessive massive bodies in the Mars-region \citep{clement18,clement21_tp,nesvorny21_tp}, and providing high-speed collisions that might produce an appropriate Mercury-analog through collisional fragmentation \citep{benz07,asphaug14,clement21_merc3}.  Our paper is organized as follows: $\S$ \ref{sect:2} summarizes a number of outstanding problems (\ref{sect:problems}) we aim to study, the success criteria used in our analyses (\ref{sect:constraints}), and our simulation setup (\ref{sect:methods}); while the results of these models are discussed in $\S$ \ref{sect:results}.

\section{Background and Methodology} 
\label{sect:2}

\subsection{Motivation}
\label{sect:problems}

With the exception of Venus' 3.4$\degr$ inclination \citep[a consequence of strong nodal coupling with Mercury via the $s_{2}$ mode:][]{nobili89,batygin15b,clement21_merc3}, both Earth and Venus' orbits are nearly circular and co-planar.  This strongly contrasts with the orbital structures of the typical Earth-Venus systems produced in many of the aforementioned formation scenarios. Even in simulations that initialize Jupiter and Saturn on circular orbits, close encounters with other large embryos during the giant impact phase can excite proto-Earth and Venus' orbits through close encounters.  Incorporating swarms of planetesimals \citep{ray06,obrien06} and collisional fragments \citep{chambers13} that provide dynamical friction to damp the orbits of growing planets can somewhat improve simulation results, however the solar system result remains at the low-excitation extreme of the distribution of model generated outcomes \citep{clement18_frag}.  After the terrestrial planets form, their eccentricities and inclinations can damp through chaotic evolution over Gyr-timescales \citep{laskar97}, however only by a factor of $\sim$2 or so.  

To illustrate the magnitude of this persistent shortcoming in the literature, we reanalyze the terrestrial analog systems of 867 simulations considering a variety of initial conditions.  Our sample of embryo accretion models \citep[simulations reported in:][]{izidoro15,clement18,clement18_frag} comprises a collection of classic extended-disk, annulus, depleted disk, instability, and fragmentation models.  Thus, the subsequent analyses broadly consider the classic \citep[e.g.:][]{chambers98}, low-mass asteroid belt \citep[e.g.:][]{izidoro15,levison15}, Grand Tack \citep[e.g.:][]{walsh11} and early instability \citep[e.g.:][]{clement18} models.  Only eight of these systems possess Earth-Venus pairs with eccentricities and inclinations within a factor of two of the real planets' (see $\langle e,i \rangle _{EV}$ in equation \ref{eqn:ei} and additional discussion below), and none acquire a value less than that of the modern planets.  The systems satisfying the former criterion included four simulations from \citet{izidoro15} considering a steep initial surface density profile of $\Sigma \propto r^{-5.5}$ for the terrestrial forming material, and four annulus models from \citet{clement18_frag} incorporating a collisional fragmentation algorithm.  Moreover, all of these marginally successful realizations were derived from integrations that modeled the giant planets on circular orbits for the entire duration of the giant impact phase.

The giant planets' acquisition of their modern orbital configuration via an epoch of mutual encounters \citep[necessary to explain the capture irregular satellites in the outer solar system:][]{nesvorny14a,nesvorny14b,deienno14} must be reconciled within all terrestrial planet formation models.  Studies considering the Nice Model's effect on the fully formed planets resoundingly conclude that it is highly unlikely that the instability did not overly excite their orbits \citep{bras09,agnorlin12,bras13} or destabilize their configuration to the point of collisions and ejections \citep{kaibcham16}.  While specific \citep{nesvorny13} instability evolutions occasionally yield adequate analogs of the modern terrestrial system if the planets begin on circular orbits \citep{roig16}, as described above, there is no clear consensus concerning the inner solar system having ever existed in such a dynamical state.  

These persistent issues motivate us to study an instability that occurs in the latter part of the $t\lesssim$100 Myr timespan after the Sun's birth \citep[consistent with multiple geochemical and small body constraints:][]{delbo17,zellner17,morb18,nesvorny18,brasser20}.  \citet{desouza21} investigated this possibility and found that the Moon-forming impact can be triggered by the instability, while only moderately over-exciting the orbits of an initially circular proto-terrestrial system consisting of the four modern planets and the Moon-forming impactor Theia.  However, the actual impact is often delayed after the instability's inception by as much as 20 Myr.  Building on these results, in the subsequent sections we model the effects of the instability on more realistic terrestrial systems produced in embryo accretion models and containing excessive numbers of planets.  Nominally, these additional bodies represent Theia, and additional Mars-analogs to be removed by the instability.  Thus, we seek to test the hypothesis that an instability at $t\gtrsim$ 50 Myr \citep{kleine09} might (a) consolidate the Venus-Earth-Theia system by triggering a Moon-forming impact with the correct geometry, (b) void unnecessary additional large embryos and planets from the Mars region, and (c) trigger fragmenting collisions that lead to the production of a small Mercury analog.

\subsection{Dynamical Constraints}
\label{sect:constraints}

The orbital architectures of numerically generated inner solar system analogs are typically constrained by the popular radial mass concentration (RMC) and angular momentum deficit (AMD) statistics \citep{laskar97,chambers01}:

\begin{equation}
        RMC = MAX\bigg(\frac{\sum_{i}m_{i}} {\sum_{i}m_{i}[\log_{10}(\frac{a}{a_{i}})]^2}\bigg)
        \label{eqn:sc}
\end{equation}

\begin{equation}
	AMD = \frac{\sum_{i}m_{i}\sqrt{a_{i}}[1 - \sqrt{(1 - e_{i}^2)}\cos{i_{i}}]} {\sum_{i}m_{i}\sqrt{a_{i}}} 
	\label{eqn:amd}
\end{equation}

While these metrics are useful for measuring the broad mass distribution and dynamical excitation of planetary systems in general, they also possess certain inherent degeneracies that can lead to the misclassification of terrestrial systems.  For instance, the RMC of a compact system of 10 low-mass planets can be exactly the same as that of a well-spaced pair of two massive bodies.  While the dynamical excitation of the inner solar system is characterized by the contrast between Earth and Venus' cold orbits and the smaller two planets' moderately excited orbits, an excessively massive Mars analog with a low eccentricity and inclination could positively weight the AMD of a numerically generated system possessing overly excited analogs of Earth and Venus.  

\begin{figure}
	\centering
	\includegraphics[width=.5\textwidth]{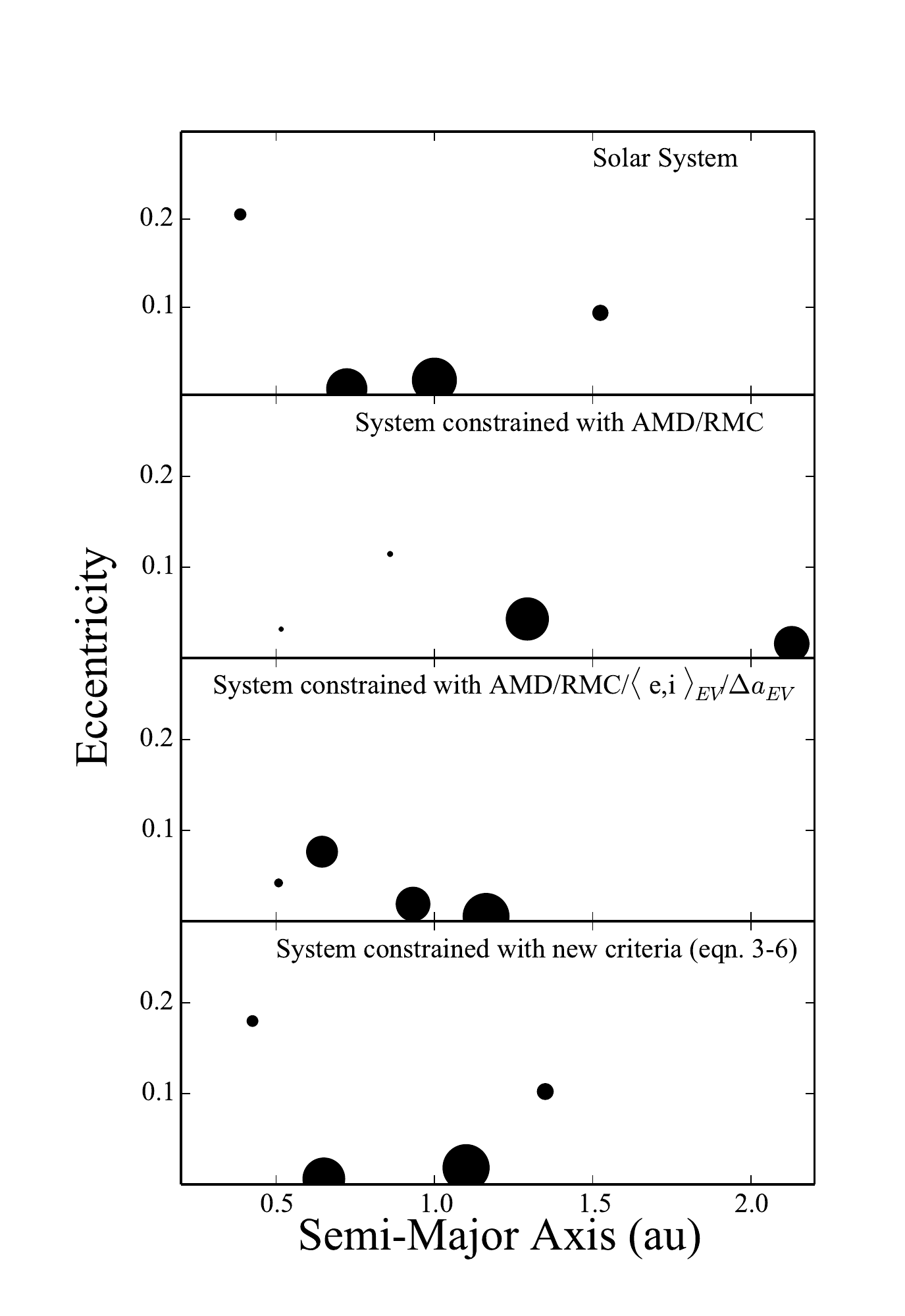}	
	\caption{Example artificial terrestrial planet systems categorized as successful by the standard AMD and RMC metrics \citep[second panel:][]{chambers01}, the modified success criteria (third panel: AMD, RMC, $\langle e,i \rangle _{EV}$ and $\Delta a_{EV}$) of \citet{nesvorny21_tp}, and the constraints defined in this work (third panel: equations \ref{eqn:ei}-\ref{eqn:mars}), compared with the modern inner solar system (top panel).}
	\label{fig:comp}
\end{figure}

To demonstrate the inadequacies of AMD and RMC, we generate 10,000 artificial inner solar system analogs utilizing our previously discussed collection of 867 late stage accretion simulations from \citet{izidoro15}, \citet{clement18} and \citet{clement18_frag}.  We take the masses, inclinations and eccentricities of all planets ($m>$ 0.05 $M_{\oplus}$ throughout this manuscript) formed as sample distributions of potential parameters for our fictitious systems.  We begin by randomly selecting one of our 867 systems, and utilizing the semi-major axes of each planet in that system as the basis for an artificial system.  We then randomly assign masses, inclinations and eccentricities from the distribution of all planets in our 867 system sample to each of these semi-major axes to create a unique system of orbits.  We also imbed a slightly modified version of the actual terrestrial system with $\sim$10$\%$ deviations randomly applied to the planets' orbital elements and masses in this set.  Only 16 of these fictitious systems satisfied the standard success criteria \citep[e.g.:][]{chambers01,obrien06,ray09a,iz14} of simultaneously possessing four planets, AMD$<$ 2 x AMD$_{SS}$ (where $SS$ stands for the solar system value) and 0.5 x RMC$_{SS}<$ RMC $<$ 1.5 RMC$_{SS}$.  Moreover, our injected version of the actual terrestrial system was ranked as the fourth best analog according to AMD and RMC (we ranked systems by averaging the percentage deviation from the solar system values for each metric).  Upon inspecting the systems deemed superior to the injected system, we find them all to be poor analogs of the actual terrestrial system for a variety of reasons.  Intriguingly, the largest planet in the system with best final AMD and RMC (plotted in the second panel of figure \ref{fig:comp}) has a semi-major axis close to that of Mars.  As the majority of the system's mass is concentrated in two large planets, the final RMC loosely resembled that of the actual terrestrial system.  Moreover, as the larger planets' eccentricities and inclinations are very small, the system's AMD is necessarily low.

Alternatives to these constraints, are rather abundant in the contemporary literature \citep[e.g.:][]{clement18,lykawka19,clement21_merc2}.  While many of these classification algorithms are rather complex, we propose a simple series of constraints that build on the recent work of \citep{nesvorny21_tp} for use in our subsequent analyses.  For all systems that form more than two planets, we define the most massive pair of neighboring planets as Earth and Venus before calculating the following parameters:

\begin{equation}
	\langle e,i \rangle _{EV} = \frac{ e_{V} + e_{E} + \sin i_{V} + \sin i_{E} }{4}
	\label{eqn:ei}
\end{equation}
\begin{equation}
	\Delta a_{EV} = a_{E} - a_{V}
	\label{eqn:a}
\end{equation}
\begin{equation}
	M_{Me} / M_{V} = \frac{\sum M_{i}(a_{i}<a_{V})}{M_{V}}
	\label{eqn:merc}
\end{equation}
\begin{equation}
	M_{Ma} / M_{E} = \frac{\sum M_{i}(a_{i}>a_{E})}{M_{E}}
	\label{eqn:mars}
\end{equation}

Here, the subscripts Me, V, E and Ma denote the respective four terrestrial planets.  The first two constraints \citep[equations \ref{eqn:ei} and \ref{eqn:a}, taken from][]{nesvorny21_tp} scrutinize Earth and Venus' dynamical excitation and radial offset, while equations \ref{eqn:merc} and \ref{eqn:mars} (new in this work) measure the total mass in planets interior to Venus and exterior of Earth.  As few dynamical studies report issues with replicating the combined mass of Earth and Venus \citep[typically the distribution of mass between the two planets is a simple consequence of the stochastic series of final giant impacts:][]{jacobson17b}, we find additional constraints on the more massive planets' masses to be unnecessary.  

We tested these new constraints using the same collection of 10,000 artificial systems.  The third panel of figure \ref{fig:comp} plots the system with final AMD, RMC, $\langle e,i \rangle _{EV}$, and $\Delta a_{EV}$ values closest to those of the solar system, and demonstrates how the two new constraints on the Earth-Venus system from \citet{nesvorny21_tp} alone are insufficient to prevent AMD and RMC from misclassifying poor analogs as successful.  However, we find that our new simple algorithm (equations \ref{eqn:ei}-\ref{eqn:mars}) effectively returns the best inner solar system analogs from our sample of 10,000 artificial systems.  In fact, we find 56 total systems, the majority of which were deemed unsuccessful by AMD and RMC, that simultaneously attain $\langle e,i \rangle _{EV}<$ 2.0 $\langle e,i \rangle _{EV,SS}$, $\Delta a_{EV}<$ 1.5 $\Delta a_{EV,SS}$, 0.0 $< M_{Me} / M_{V} \leq$ 0.2 and 0.0 $< M_{Ma} / M_{E} \leq$ 0.3.  Moreover, this classification scheme successful ranks our injected inner solar system analog as the best system (plotted in the bottom panel of figure \ref{fig:comp}).  

This result demonstrates how the AMD and RMC metrics can potentially misclassify successful simulations as unsuccessful, and exaggerate the merits of a non-viable simulation batch.  For instance, poor solar system analogs similar to that middle panel of figure \ref{fig:comp} might boost the rate of simulations with adequate AMD values.  We do not contend that previous studies relying on these metrics overlooked successful simulations.  Indeed, most papers tend to employ complex classification algorithms to separate good systems from bad ones \citep[e.g.:][]{clement18,lykawka19}.  Moreover, we do not plead to diminish the importance and efficacy of other standard constraints related to the planets' formation timescales \citep[e.g.:][]{Dauphas11}, Earth's water mass fraction \citep[e.g.:][]{ray07}, small body distributions \citep[e.g.:][]{obrien07} and late material delivery \citep[i.e.: the late veneer:][]{bottke10,raymond13}.

We utilize these new constraints in our subsequent analyses of the effects of a late Nice Model instability on the Earth-Venus-Theia system.  Given the disparate initial conditions utilized in our models, we let the pre-instability classification of each body be ambiguous.  Thus, Theia is not defined as the n-th body from the Sun, but rather the smaller body involved in the first collision between two proto-planets.  We then define Earth and Venus according to their distance from the Sun as described above in simulations that retain exactly two large planets.  An obvious drawback of this methodology is the potential for neglecting massive objects surviving in between the orbits of Earth and Venus.  We note no instances of $\sim$Mars-mass embryos attaining such orbits, and discuss the cases that retain all three massive proto-planets independently.  Other instances of excessive planets existing in a system (e.g.: interior to Mercury, between Mercury and Venus, between Earth and Mars, or exterior to Mars) would be reflected in our $M_{Me} / M_{V}$ and $M_{Ma} / M_{V}$ statistics.

\subsection{Numerical Simulations}
\label{sect:methods}

\subsubsection{Initial condition selection}
\begin{figure}
	\centering
	\includegraphics[width=.5\textwidth]{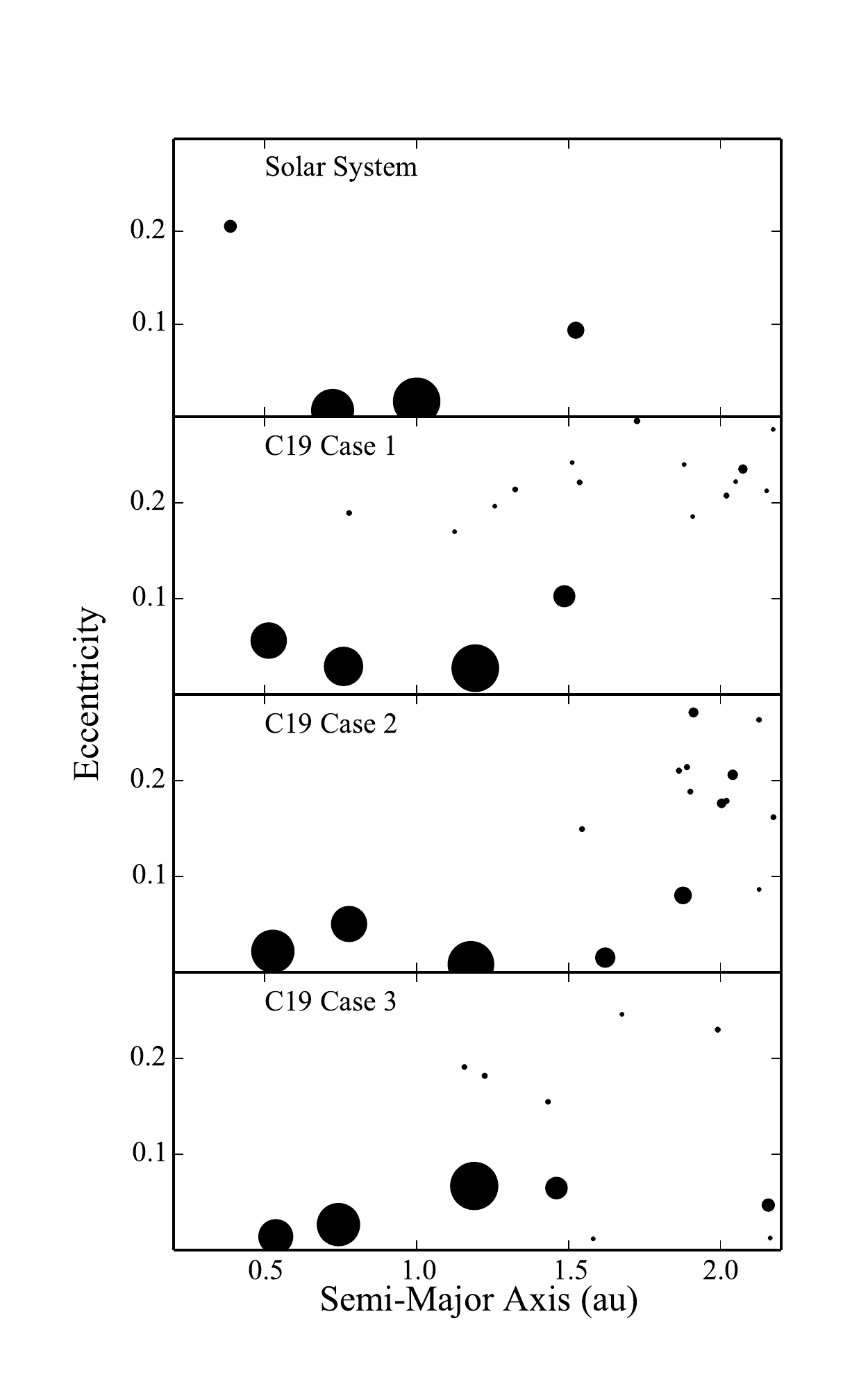}
	\caption{Pre-instability configurations of the inner solar system utilized in our study \citep[note that these systems are derived directly from outputs of simulations of the classic model of terrestrial planet formation incorporating a collisional fragmentation model from][]{clement18_frag}.  The size of each point is proportional to the object's mass.  An inclination version of this plot is provided in figure \ref{fig:ics_inc}.}
	\label{fig:ics}
\end{figure}

We begin by selecting three inner solar system analogs from \citet{clement18_frag} to use as initial conditions for our present study (cases 1, 2 and 3 in figures \ref{fig:ics} and \ref{fig:ics_inc}).  In order to  study the effects of a Nice Model instability occurring at near the end of the giant impact phase, we utilize the state of each system 5 Myr after the final embryo impact on an Earth, Venus or Theia analog (50.5, 68.2, and 147.0 Myr for cases 1, 2 and 3, respectively; see additional discussion below).  More precisely, we integrated these systems to $t=$ 500 Myr in previous work, and utilized the coordinate outputs from the models at the aforementioned times as initial conditions for our current study.  In the original simulations reported in \citet{clement18_frag}, these systems were evolved in the presence of a static Jupiter and Saturn, and thus were not perturbed by the Nice Model instability.  Each system was generated from an initial disk of 100 embryos and 1,000 planetesimals extending from 0.5-4.0 au.  The objects were assigned near-circular orbits at the beginning of the simulation, and semi-major axes such that the disk's surface density profile was proportional to $r^{-3/2}$.  The integration time-step was 6 days, and Jupiter and Saturn were included in the models on their presumed pre-instability orbits \citep[i.e.: in a 3:2 mean motion resonance, see:][]{deienno17}.  Each disk was integrated for 200 Myr utilizing a modified version of the \textit{Mercury6} Hybrid integrator \citep{chambers99} that incorporates algorithms designed to account for the effects of collisional fragmentation.  The minimum mass of ejected collisional fragments in our model is set to 0.005 $M_{\oplus}$ \citep{chambers13,wallace17}.

The rationale behind our selection of these specific systems is threefold.  First, we look for systems possessing exactly three planets with $a<$ 1.3 au and $m>$ 0.1 $M_{\oplus}$ that are nominally intended to represent Earth, Venus and Theia \citep[e.g.:][]{asphaug14_moon}.  Next, we discard any systems with bodies more massive than 0.2 $M_{\oplus}$ in the Mars-region $a>$ 1.3 au.  This limits our potential Mars analog masses to within a factor of $\sim$2 of Mars' modern mass, which we assess to be a reasonable compromise given the goals of our investigation.  Finally, we select the three systems with the lowest Venus-Earth-Theia system AMD.  As discussed in $\S$ \ref{sect:problems}, each system's AMD and $\langle e,i \rangle _{EV}$ values are larger than those of the solar system.  The fact that none of our initial Earth and Venus analogs have eccentricities and inclinations less than those of the real planets is an important limitation of our study, and also distinguishes our work from that of \citet{desouza21}.  Thus, for our scenario to be successful the process of the instability triggering the Moon-forming impact (and the removal of Theia from the system) must damp the orbits of Earth and Venus.

While not the central focus of our current study, we also verify that the pre-impact orbit of Theia in our systems is consistent with the parameters required to give an ideal Moon-forming impact as calculated in \citet{jackson18}.  It is important to note that each of our tested cases explicitly attempt to generate a Moon-forming impact between two roughly equal-mass bodies as proposed by \citet{canup12}, as opposed to the scenario where Theia's mass is roughly similar to that of Mars as suggested in classic studies \citep[e.g.:][]{cameron76,benz86,canup04}.  Though we mostly leave the study of this setup for future work, we briefly analyze the collisional geometries of impacts between additional embryos in the Mars-region (figures \ref{fig:ics} and \ref{fig:ics_inc}) and the outermost massive proto-planet (Earth analog).

While our instability simulations consider the state of embryo accretion models at $t=$ 51, 68 and 147 Myr for cases 1, 2 and 3 as initial conditions, we verified that each system attained its quasi-stable configuration of proto-planets within the first 50 Myr of the initial accretion simulation.  Specifically, we confirmed that no additional objects larger than Mars impacted Earth, Venus or Theia after this point.  Moreover, we performed an extended 500 Myr N-body integration of each system to confirm that the massive bodies ($m>$ 0.1 $M_{\oplus}$) are stable in the absence of a giant planet instability.  Given this degree of stability, we argue that the precise timing of the instability in our models can be considered mostly arbitrary.  Thus, our simulations are intended to serve as a reasonable proxy for studying the effects of a Nice Model that transpires between $\sim$50 and 100 Myr after gas dispersal \citep[consistent with dynamical and geophysical constraints:][]{nesvorny18,morb18,brasser20,ribeiro20,nesvorny21_e_nep} on a nearly formed terrestrial system grown from an extended disk of planet forming material \citep{chambers98}, rather than a concentrated region of planetesimals around 1 au \citep{hansen09,iz14,izidoro21_nat}.  However, systems similar to our cases 1, 2 and 3 also emerge from numerical simulations of terrestrial planet formation invoking alternative disk initial conditions \citep[see, for example:][]{walsh16,deienno19,walsh19}.

\subsubsection{Instability models}
We utilize the same numerical pipeline described in detail in \citet{clement21_tp} to generate a significant number of distinctive instability evolutions that successfully replicate the modern outer solar system in broad strokes.  Essentially, our approach is to continuously restart instability simulations until the evolution adequately reproduces Jupiter's moderately excited fifth eccentric eigenmode, and its orbital period ratio with Saturn \citep[both properties are key drivers of dynamical evolution in the inner solar system:][]{bras09,morby10,clement21_instb}.  Specifically, we require instabilities' excite $e_{J}$ to greater than 0.03 and drive $P_{S}/P_{J}$ to between 2.3 and 2.5 within 200 kyr.  We also terminate simulations where $e_{J}$ damps to below 0.015, or those that exceed $P_{S}/P_{J}=$ 2.8 at any point.  When a successful outer solar system is attained, the integration is continued up to $t=$ 100 Myr.  As in \citet{clement21_tp}, our final sample of fully evolved systems include a mix of five giant planet instability evolutions beginning from 3:2,3:2,3:2,3:2 \citep{nesvorny12} and 2:1,4:3,3:2,3:2 \citep{clement21_instb} resonant chains, however we note no significant differences between the final properties of these two sets of systems in our subsequent analyses.  Each of these simulations utilize the same fragmentation version of the \textit{Mercury6} hybrid integrator described above with a 6.0 day time-step \citep{chambers99,chambers13} and an identical minimum fragment mass setting.  We remove objects that attain heliocentric distances in excess of 1,000 au or make perihelia passages less than 0.1 au.  Through this process, we attain a sample of 14, 21 and 21 systems for cases 1, 2 and 3, respectfully.

\section{Results}
\label{sect:results}
`
\subsection{Consolidating the Venus-Earth-Theia system}
\subsubsection{Equal-mass impact model}
\label{sect:moon}

Of our 56 simulations that obtained successful instabilities and were evolved for 100 Myr, 36 yield at least one giant impact between a pair of large proto-planets (Earth, Venus or Theia).  We find that such giant impacts occur with roughly equal likelihood in our three cases (figure \ref{fig:ics}), although case 1 yields the lowest fraction of giant impacts (6/14 total simulations).  As described in \citet{desouza21}, this is likely the result of the larger initial orbital spacing between proto-planets in case 1  ($\sim$0.25 and 0.45 au between each pair of embryos as compared to $\sim$0.2 and 0.4 au for cases 2 and 3).  Of the remaining systems that do not yield potential Moon-forming impacts after being evolved through the giant planet instability, 11 retain all three of the most massive terrestrial bodies, and the remaining systems lose terrestrial planets via collisions with the central body.

While these results are encouraging in demonstrating the instability to be an efficient trigger of the Moon-forming impact \citep[consistent with the findings of][]{desouza21}, we observe a diverse spectrum of possible evolutionary histories for the massive bodies Earth, Venus and Theia.  In some simulations the three proto-planets eventually combine to form a single massive terrestrial object, while others yield a satisfactory giant impact between two of the bodies before losing the third to a collision with the Sun.  In this manner, out of 6 total case 1 systems that produced giant impacts, only finish with exactly two analogs of Earth and Venus.  Similarly, 12/16 case 2 simulations are successful in the manner, compared to just 7/14 for case 3.  Moreover, many of these initial giant impacts involve the innermost planet (nominally the system's Venus analog).  Specifically, 28 of our 36 giant impact producing simulations exhibit an impact on Venus after the instability ensues.  Excessive collisions involving Venus were also observed in \citet{desouza21}, and are likely related to the shorter dynamical timescales at Venus' orbit, along with strong eccentric coupling between the $g_{2}$ and $g_{5}$ frequencies during the instability and subsequent Jovian residual migration phase \citep{bras09}.  As 5 of 16 case 2 Venus analogs do not experience a giant impact\footnote{We note that none of our simulations where a single giant impact occurs involving the second and third massive bodies, Earth and Theia, also retain exactly one Mars analog (see additional discussion in $\S$ \ref{sect:mars}}, and this simulation batch also yields the largest fraction of total systems producing exactly two Earth and Venus analogs, we assess it to be the most successful set of initial conditions in our present study.  

\begin{figure}
\centering
\includegraphics[width=.50\textwidth]{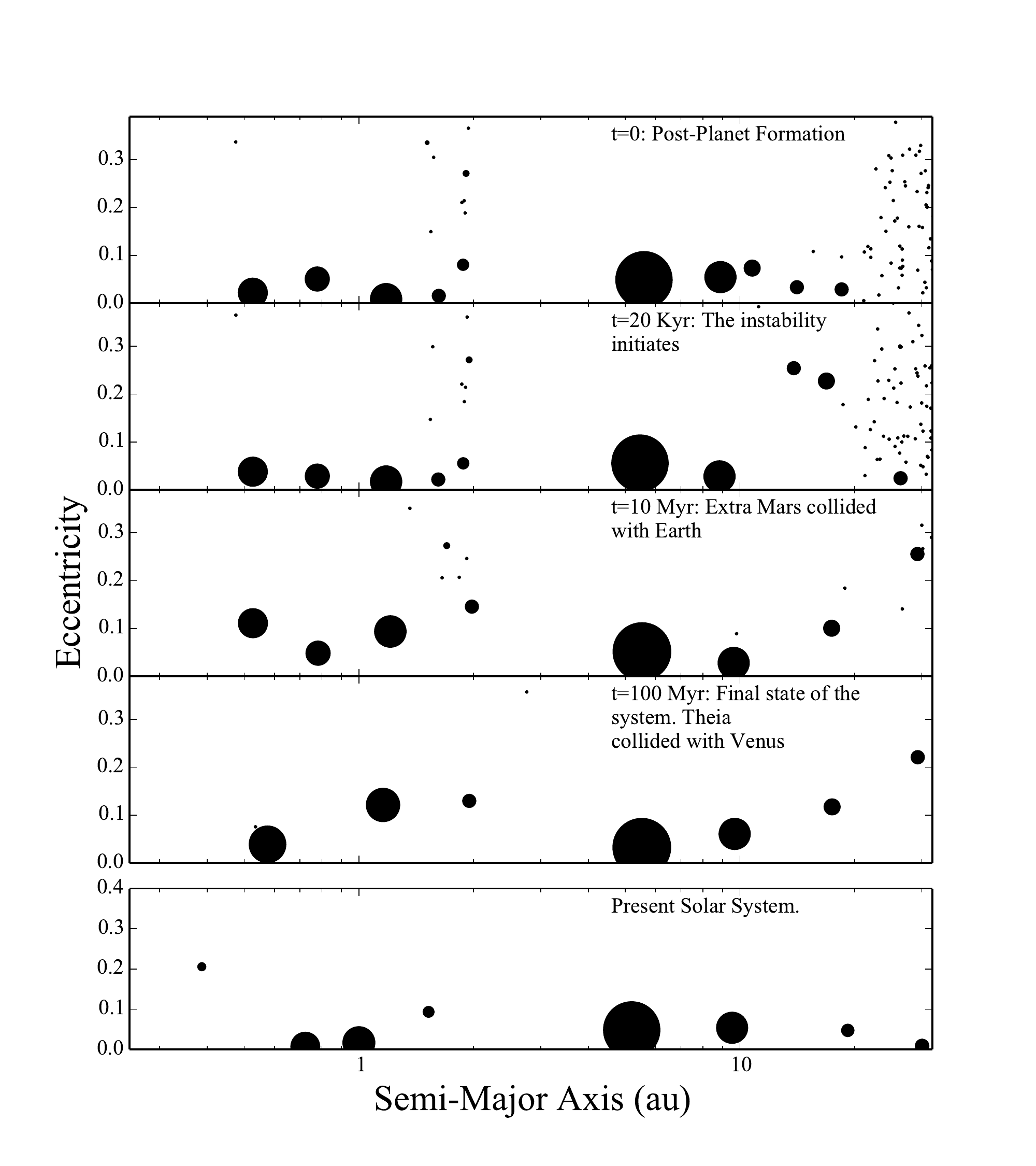}
\caption{Example evolution of a successful extended instability simulation.  The initial inner solar system is derived from a simulation in \citet[][incorporating the effects of collisional fragmentation]{clement18_frag} that did not include an instability model and formed three planets in the Earth/Venus region, two Mars-like planets in the Mars region, and stranded two additional un-accreted embryos in the vicinity of Mars. The size of each point corresponds to the mass of the particle.  The final terrestrial planet masses are $M_{Venus}=$ 1.32 ($M_{Venus,SS}=$ 0.815), $M_{Earth}=$ 1.09 and $M_{Mars}=$ 0.15 ($M_{Mars,SS}=$ 0.107) $M_{\oplus}$.  This simulation also satisfies the four outer solar system success criteria ($e_{55}=$ 0.036, $P_{S}/P_{J}=$ 2.33; compared with $e_{55,ss}=$ 0.044, $P_{S}/P_{J,ss}=$ 2.49) described \citet{clement21_instb}.  We note that, while the eccentricities of Uranus and Neptune are larger than in the actual solar system, the effect of this over-excitation on the inner solar system is negligible \citep{bras09}.  However, such an evolution would not be successful in matching certain constraints from the Kuiper Belt \citep[e.g.:][]{nesvorny15a,nesvorny21_e_nep}.}
\label{fig:time_lapse}
\end{figure}

An example evolution from our case 2 simulation batch is plotted in figure \ref{fig:time_lapse}.  In this model, the system finishes the initial terrestrial planet formation simulation reported in \citet{clement18_frag} with three large planets in the Earth-Venus region, three Mercury-Mars-mass planets in the Mars region, and a distribution of collisionally-generated debris permeating the entire inner solar system.  As a result of perturbations from the Nice Model instability, Venus collides with and accretes the additional large body, one of the unnecessary Mars analogs merges with Earth, and the remaining excessive bodies in the Mars-region attain hyperbolic orbits and are ejected.

\begin{figure}
	\centering
	\includegraphics[width=.5\textwidth]{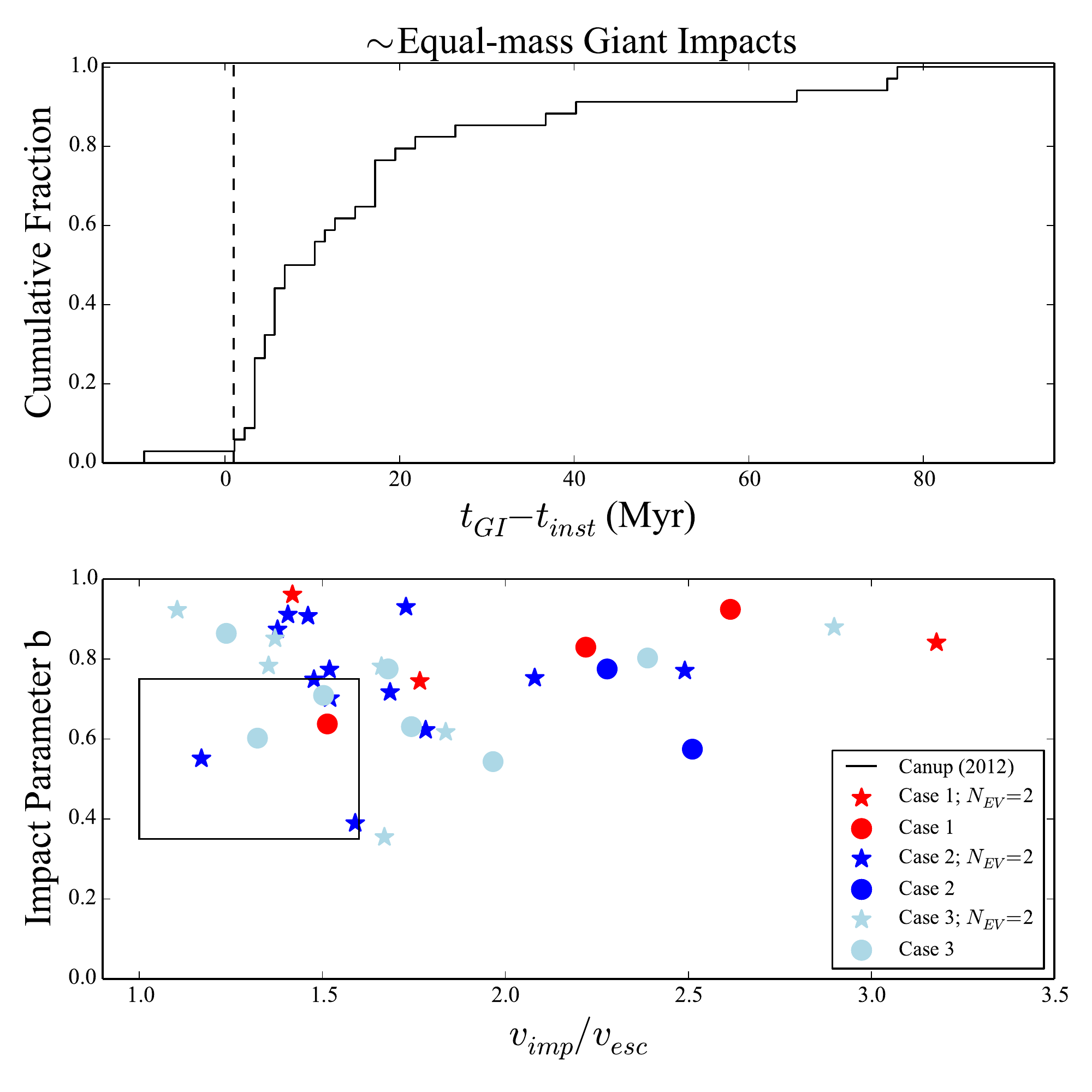}
	\caption{Top panel: Cumulative fraction of initial giant impact (GI) time for any Earth, Venus or Theia analog in our various simulation sets with respect to the instability time (dashed line: determined by a ``jump'' in Jupiter's semi-major axis).   Bottom panel: Distribution of impact parameters (b) and impact velocities (scaled by the two-body escape speed: $v_{imp}/v_{esc}$) for these first giant impacts.  The color of each point corresponds to the simulation case (red, blue and light blue for cases 1, 2 and 3, respectively: figure \ref{fig:ics}).  Stars correspond to simulations that finish with exactly two planets, Earth and Venus analogs (i.e.: successfully transform a system of three large proto-planets into two).  The box roughly denotes the preferred impact parameters for the \citet{canup12} Moon-formation model.}
	\label{fig:moon}
\end{figure}

The top panel of figure \ref{fig:moon} plots the cumulative distribution of giant impact times measured with respect to the instability time ($t_{GI}-t_{inst}$, where GI stands for Giant Impact).  Consistent with \citet{desouza21}, we observe that the first impact between two of the three massive proto-planets in a system is often delayed by  $\sim$5-20 Myr.  However, our systems destabilize much quicker (median $t_{GI}-t_{inst}=$ 7.4 Myr) than those of \citet{desouza21} (median $t_{GI}-t_{inst}$ in excess of 20 Myr for the majority of cases tested).  This is likely the result of the higher initial eccentricities (figure \ref{fig:ics}) considered in our work.  We also looked for correlations between the giant planets' final orbital configuration and the distribution of $t_{GI}$ times.  As our simulation pipeline necessarily selects for evolutions that reasonably replicate Jupiter and Saturn's modern dynamical configuration, our sample of final outcomes did not possess a sufficiently broad spread of $e_{J}$, $e_{S}$ and $P_{S}/P_{J}$ values to reveal any meaningful trends.  Thus, future work studying a broader spectrum of instability outcomes \citep[e.g.:][]{clement18_ab} with a larger number of simulations will be valuable for determining how well $t_{inst}$ might be able to be constrained by $t_{GI}$.

Certain isotopic constraints such as $^{182}$Hf - $^{182}$W systematics have been interpreted to date the Moon-forming impact's occurrence at around $t\simeq$ 50-100 Myr \citep[e.g.:][]{kleine09}.  As the majority of our systems experience a giant impact within the first 10 Myr after the instability occurs, and our systems are representative of a plausible dynamical state of the inner solar system between $t=$ 50 and 100 Myr, our scenario is potentially consistent with the results of many recent studies favoring a Nice Model disturbance in the first 100 Myr after the solar system's birth \citep[see $\S$ \ref{sect:1} and][]{morb18,nesvorny18,ribeiro20,brasser20}.

The bottom panel of figure \ref{fig:moon} plots the impact parameter, b, and relative collision velocity (scaled by the two-body escape velocity; $v_{imp}/v_{esc}$) for each initial giant impact involving two of the three massive proto-planets (Earth, Venus and Theia) in our simulations.  Points corresponding to systems that complete the integration with exactly two analogs of Earth and Venus are depicted with stars.  It is clear from the distribution of these points that no clear preference towards higher or lower collision speeds exists for the Earth-Venus analog systems.  While a complete study of the consistency between the impacts observed in our simulations and the equal-mass impact model of \citet{canup12} is beyond the scope of this work, we plot the approximate range of preferred parameters from that work with a black box in figure \ref{fig:moon} for comparison\footnote{It is important to note that multiple factors are involved in constraining an ideal Moon-forming impact \citep[see][for a review]{canup14_rev}.  Among others, the replication of the modern Earth-Moon system angular momentum, the pairs' mass ratio and disparate iron contents \citep{reufer12,cuk12,lock18}, and the accretion history that determines Theia's isotopic composition \citep{kaibcowan15,quarles15} are supremely important.}.  It is clear that a reasonable number of our successful simulations yield the characteristic low-velocity, grazing impact geometry preferred in the analysis of \citet{canup12}.

\subsubsection{Canonical Mars-mass impactor model}

Since we explicitly choose initial terrestrial systems containing exactly three massive proto-planets, our models are not intended to serve as hypothetical initial conditions for the canonical Moon-forming impact model \citep{chapman75,benz86,canup04} where a $\sim$Mars-mass projectile collides with the almost fully grown proto-Earth.  Nevertheless, we briefly explore the possibility of a smaller projectile forming the Moon (nominally one of the additional large embryos in the Mars-region) in our scenario with figure \ref{fig:mars}.  In total, 13 of our Earth analogs (outermost of three large proto-planets for each case depicted in figure \ref{fig:ics}) collide with a $\sim$Mars-mass object (defined here as 0.025 $<m<$ 0.3 $M_{\oplus}$) during the Nice Model evolution.  Thus, in spite of our efforts to select initial terrestrial configurations that might produce a collision between two $\sim$0.5 $M_{\oplus}$ objects exterior to Venus' orbit, such events (8 occurrences) are outnumbered by Mars-mass potential Moon-forming impactors originating in the Mars-region.  While the cumulative distributions of giant impact times with respect to the instability time (top panel of figure \ref{fig:mars}) is similar to that of the equal-mass giant impacts depicted in figure \ref{fig:moon}, the impacts themselves tend to occur at lower relative velocities and lower impact angles (bottom panel of figure \ref{fig:mars}).  This is consistent with the preferred impact geometries found in hydrodynamical simulations of the canonical Moon-forming impact \citep{canup04,asphaug14_moon}.  However, we do observe three impacts that occur at $v_{imp}/v_{esc}>$ 2.0.  Notably, such events are consistent with the modified version of the classic model proposed in \citet{cuk12} that invoke angular momentum loss via the evection resonance with the Sun from a primordially faster-spinning Earth-Moon system.  In fact, one such case 3 simulation provides an excellent match to the favored simulation from that work (light blue point inside of the dashed-line box in figure \ref{fig:mars}).

\subsection{Final system dynamics}

\begin{figure}
	\centering
	\includegraphics[width=.48\textwidth]{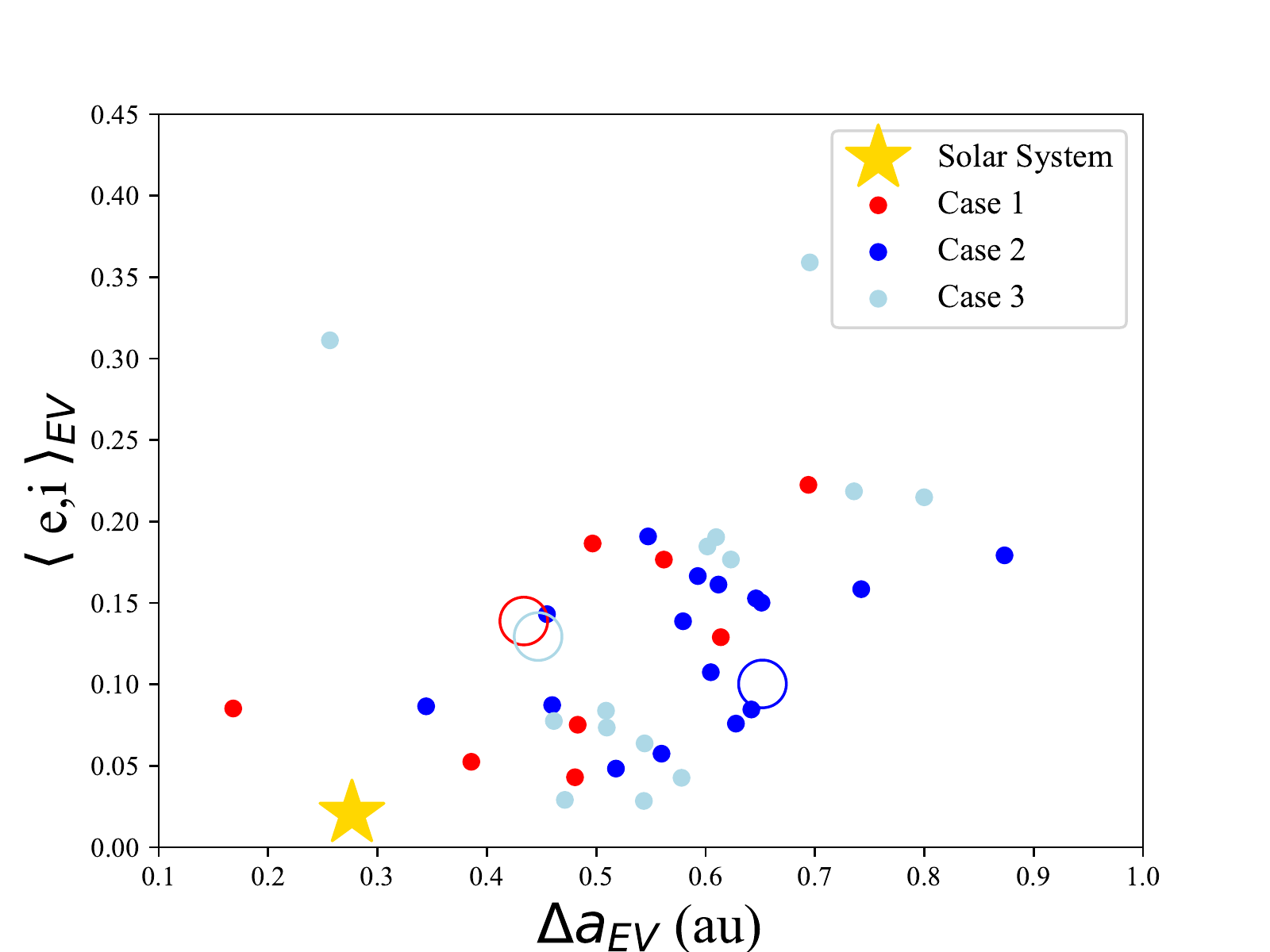}
	\qquad
	\includegraphics[width=.48\textwidth]{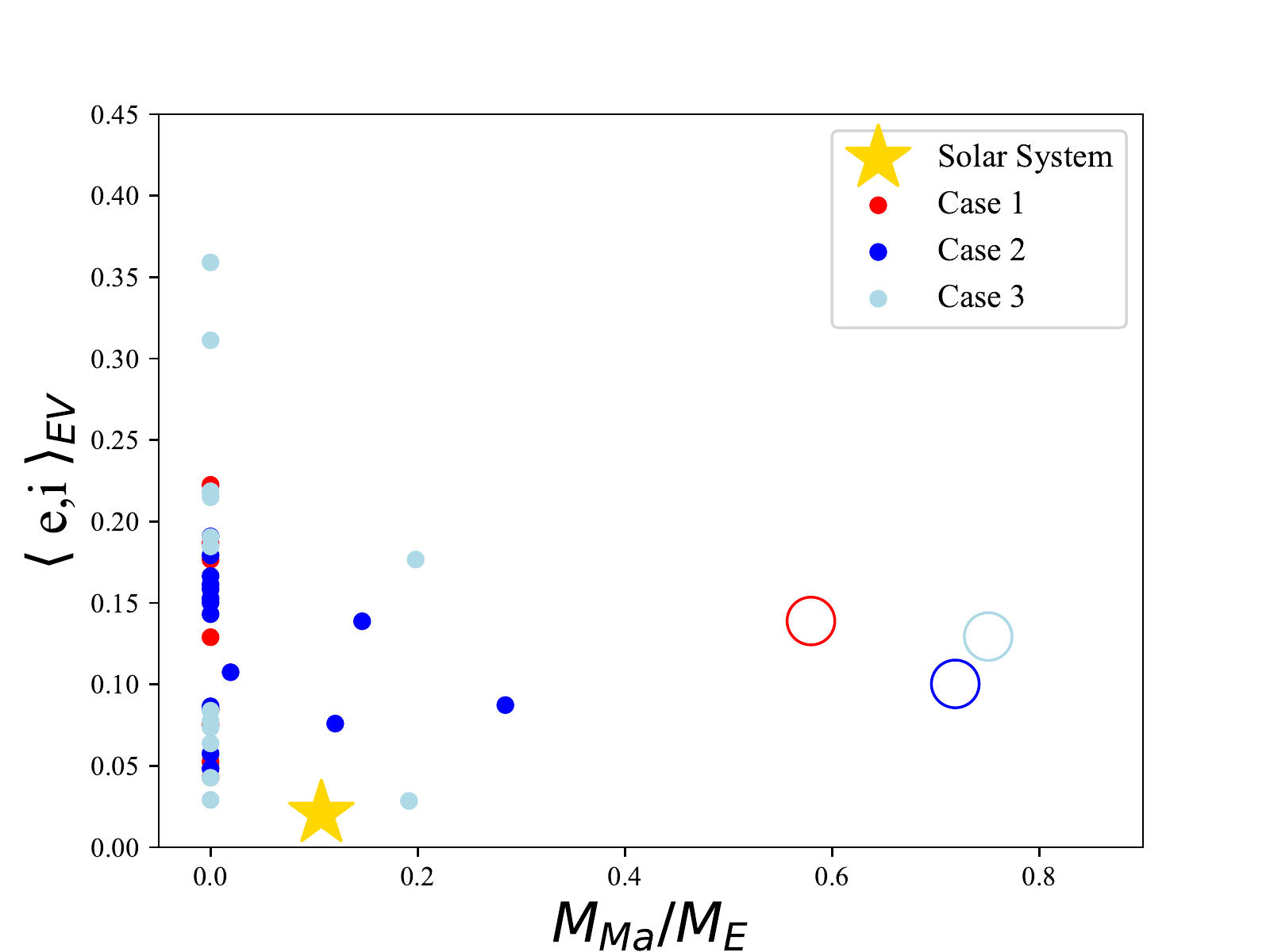}
	\qquad
	\includegraphics[width=.48\textwidth]{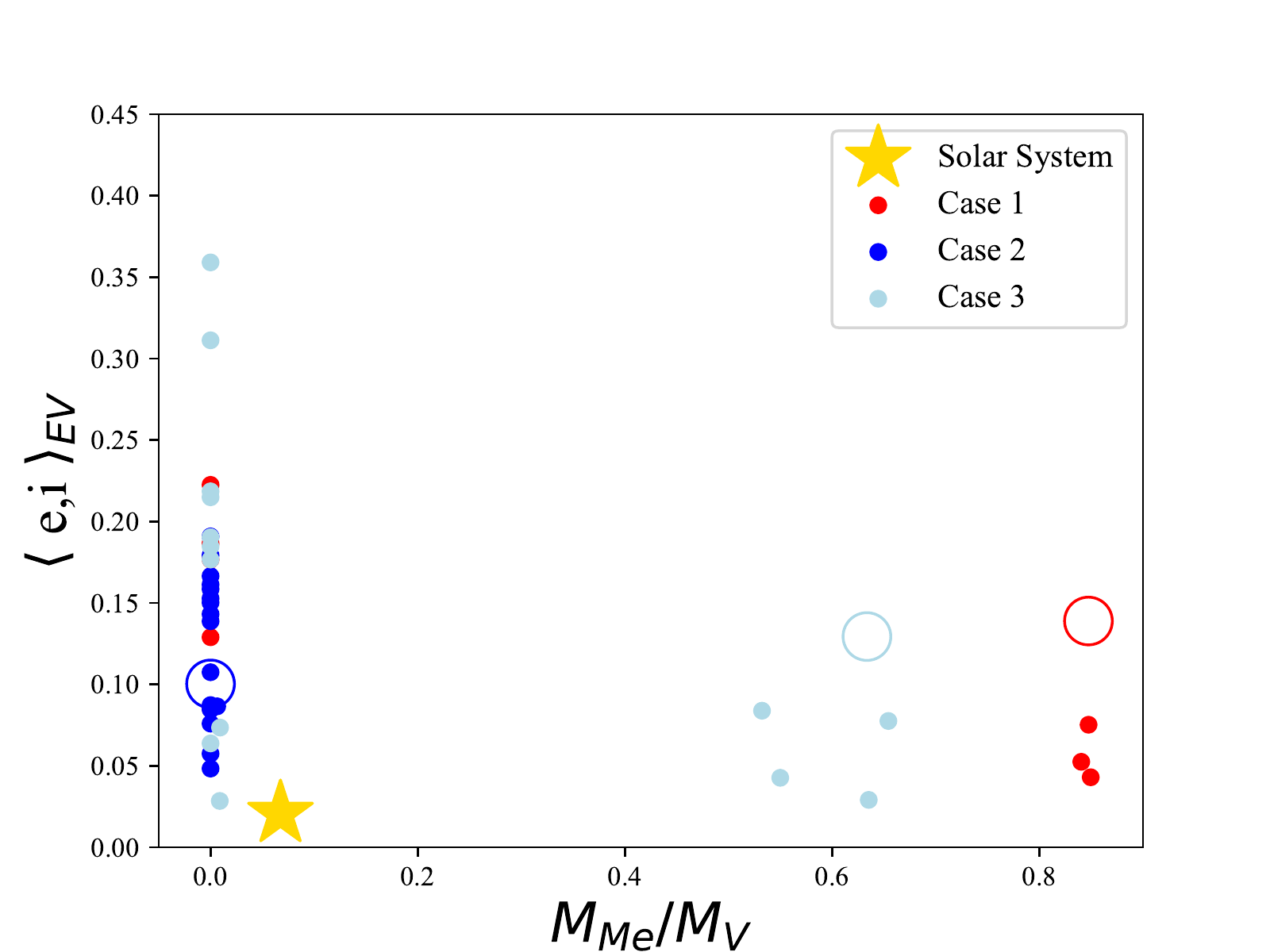}
	\caption{Distribution of final system properties (points) compared to the initial initial value for each case (open circles; see figure \ref{fig:ics}) and the modern solar system value.  The different point colors red, blue and cyan denote simulations from our three respective cases.  The top panel plots the time-averaged $\langle e,i \rangle _{EV}$ against $\Delta a_{EV}$, while the second and third panels respectively compare $M_{Ma} / M_{V}$ and $M_{Me} / M_{E}$ with Earth and Venus' excitation, $\langle e,i \rangle _{EV}$ (see equations \ref{eqn:ei}-\ref{eqn:mars}).  Points along the left vertical axes in the bottom two panels belong to systems that form no Mercury or Mars analog.}
	\label{fig:results}
\end{figure}

In this section we scrutinize the final dynamical structures of our post-instability systems utilizing the constraints described in $\S$ \ref{sect:constraints}.  Figure \ref{fig:results} summarizes these results, and clearly elucidates the fragility of our systems during the epoch of giant planet migration \citep{kaibcham16}.  Specifically, while the solar system values of $\Delta a_{EV}$, $M_{Ma}/M_{E}$ and $M_{Me}/M_{V}$ are within our spectrum of simulated outcomes, consistent with previous studies of terrestrial accretion utilizing similar methodologies \citep[see the discussion in $\S$ \ref{sect:problems} concerning the results of:][]{izidoro15,clement18,clement18_frag}, none of our systems adequately replicate Earth and Venus' dynamically cold orbits.  The subsequent three sections discuss this, and two other important shortcomings of our modeled scenario.

\subsubsection{Replicating Earth and Venus' orbital configuration}

The top panel of figure \ref{fig:results} demonstrates the challenges involved with replicating the precise Earth-Venus system in our models.  It is important to note that this analysis considers all of our simulations, including those that do not form adequate Mars analogs.  Though many of our evolved systems are poor inner solar system analogs in this sense, it is encouraging that some systems actually improve after being destabilized by the Nice Model instability (see figure \ref{fig:results}).  This is the case for approximately one third (18/56) of our systems, and is the result of angular momentum transfers between proto-planets prior to the loss of excessive bodies or collisional fragments by ejection merger with the Sun.  While the majority of our systems finish with excessively large values of $\Delta a_{EV}$, the actual value lies within the spectrum of outcomes derived from our runs.  We note that most of the systems finishing with $\Delta a_{EV} \gtrsim$ 0.5 au experience an initial giant impact involving Venus (note that this threshold is different than 1.5 x $\Delta a_{EV}$ discussed earlier in the text).  Indeed, the three closest Earth-Venus orbital spacings produced in our study (left-most red, light blue and blue points in the top panel of figure \ref{fig:results}) all occur in systems where the outer two massive proto-planets collide, and the final system includes exactly two Earth and Venus analogs.

While a small number of our simulations finish with satisfactory values of $\Delta a_{EV}$, consistent with previous studies \citep[e.g.:][]{chambers01,ray09a,izidoro15,clement18_frag}, values of $\langle e,i \rangle _{EV}$ similar to the solar system's low degree of excitation are very rare among our systems (vertical axis of all panels in figure \ref{fig:results}), and no system finishes with $\langle e,i \rangle _{EV} < \langle e,i \rangle _{EV,SS}=$ 0.021.  The median final values of $\langle e,i \rangle _{EV}$ in our simulation sets are 0.11, 0.14 and 0.13 for cases 1, 2 and 3, respectufully.  While this is clearly unsatisfactory, it is important to note that these results are not particularly different from those derived from simulations that do not incorporate a giant planet instability model, or those that consider an instability at $t \lesssim$ 100 Myr.  Indeed, the median $\langle e,i \rangle _{EV}$ values in our collection of past simulations from the literature that consider a static outer solar system are 0.11 for the depleted disk models of \citet{izidoro15}, 0.10 for the annulus models from \citet{clement18_frag}, and 0.15 for the classic extended disk simulations from that same paper.  Similarly, our collection of instability simulations from \citet{clement18_frag} have median $\langle e,i \rangle _{EV}=$ 0.15.  \citet{nesvorny21_tp} reported four outcomes with $\langle e,i \rangle _{EV} < \langle e,i \rangle _{EV,SS}$ in all instability cases tested (their figure 13), however each of these results were obtained in instabilities where Jupiter's final eccentricity was under-excited compared to the modern value ($e_{55}=$ 0.025 compared to 0.044 in the modern solar system).  As the eccentricity excitation in the inner solar system acquired during the $g_{5}$ sweeping phase \citep{bras09} of the instability is likely lower in these systems than in the actual solar system, it is possible that these low final $\langle e,i \rangle _{EV}$ values are a result of Jupiter's under-excitation.  Nevertheless, the results of \citet{nesvorny21_tp} represent perhaps the best matches to Earth and Venus' eccentricities in any model from the literature that accounts for the Nice Model instability.  Given the poor analogs of Earth and Venus produced in our work, we conclude that it is more likely that the terrestrial system's dynamical state was different at the time of the instability's onset.

\subsubsection{Removing unnecessary additional Mars analogs}
\label{sect:mars}

The middle panel of figure \ref{fig:results} compares the values of $M_{Ma}/M_{E}$ before (stars) and after (points) our simulated instabilities.  When interpreting this plot, it is important to note that, for our initial case 2 configuration, our classification algorithm classifies the outer two large proto-planets as Earth and Venus analogs (referred to as Earth and Theia in the discussion of the Moon-forming impact in $\S$ \ref{sect:moon}).  Contrarily, in cases 1 and 3 the inner two objects (Venus and Theia in the vernacular of the previous section) are considered Earth and Venus analogs. However, this initial classification has no bearing on the subsequent system classification.  Regardless of initial classification, all giant planet instability simulations are re-classified in accordance with our classification algorithm ($\S$ \ref{sect:constraints}).

Consistent with the findings of \citet{clement18}, it is clear that, in all cases the instability efficiently removes mass from the Mars-forming region.  Moreover, there is no clear connection between the excitation or over-excitation of the Earth-Venus system and the final mass of the Mars analog.  However, in all but six instances, this process is too efficient, and all material is removed from the Mars-region.  This issue of forming no Mars analog was also noted in \citet{clement18}, and is a reason why the authors preferred an instability at $t\lesssim$ 10 Myr; when sufficient planetesimal mass was available to damp the instability-induced eccentricity excitation in the Mars-region and save potential Mars-analogs from loss.  While some of our simulations do indeed perturb the Mars-region in a manner such that they finish with a exactly one Mars-analog interior to the Earth-Venus system, the rarity of such outcomes coupled with the low probability of producing a giant impact on Earth rather than Venus does strongly speak against our modeled scenario.  Of our sample of 56 instability simulations, only two (both from case 2) produce giant impacts on the outermost large proto-planet (Earth analog), and simultaneously finish with analogs of Earth, Venus and Mars (i.e.: $N_{EV}=$ 2 and $M_{Ma}/M_{E}<$ 0.3; $\ne$ 0.0).  In both such cases (one of which is plotted in figure \ref{fig:time_lapse}), an additional Mars analog collides with Earth (figure \ref{fig:mars}), Venus merges with Theia, and a single Mars analog survives.  However, both of these systems have higher than average final $\langle e,i \rangle _{EV}$ values.  We conclude that the challenges involved with retaining Mars analogs in our scenario are probably applicable to all terrestrial planet formation models, and strongly suggest that the instability occurred earlier than $t=$ 50-100 Myr.

\subsubsection{Generating Mercury as a collisional fragment}

While few of our simulations yield non-zero values of $M_{Me}/M_{V}$ (bottom panel of figure \ref{fig:results}), fragment producing collisions on Venus \citep[a potential origin for Mercury:][]{asphaug14} are common in our models.  Indeed, 11 of the 28 giant impacts involving the inner two large proto-planets (nominally Venus and Theia) produce collisional fragments \citep{leinhardt12}.  Although the majority of these collisional fragments merge with the Sun or are quickly re-accreted by Venus, two systems in our case 3 batch retain a collisional fragment on an orbit interior to Venus'.  While both of these Mercury analogs are on orbits that are too close to Venus', it is important to note the somewhat arbitrary prescription for fragment creation and ejection utilized in our code.  When a fragmenting collision is detected, the total remnant mass is calculated using relations from \citep{leinhardt12}.  This mass is divided into a number of equal-mass fragments that are ejected at $v \simeq$ 1.05 $v_{esc}$ at uniformly spaced directions in the collisional plane.  If the ejection velocity or mass distribution between fragments were different, it is certainly plausible that a Mercury-forming impact similar to those modeled in \citet{asphaug14} could occur in our scenario.  However, as discussed above, such a late impact on Venus is potentially problematic.

\section{Conclusions}

We simulated the effects of the Nice Model instability on three configurations of inner solar system material that represent plausible states of the terrestrial system around 50-100 Myr after the dissipation of the gaseous disk.  This timing of the giant planet instability is consistent with that inferred from a number of geophysical, geochemical and dynamical constraints \citep[e.g.:][see additional discussion in $\S$ \ref{sect:1}]{morb18,nesvorny18,brasser20}.  Consistent with the recent work of \citet{desouza21}, we find that the instability has the potential to trigger the giant impact that formed the Moon.  As such giant impacts consistently occur within 5-10 Myr after the jump in Jupiter's semi-major axis, the timing of the Moon-forming impact inferred via isotopic dating \citep[e.g.:][]{kleine09,rudge10} provides an interesting constraint on the instability's timing in our scenario.  Thus, our work demonstrates that stable configurations of five or more proto-planets regularly produced in embryo accretion models invoking a variety of different initial conditions \citep[e.g.:][]{walsh16,deienno19,clement18_frag} can be destabilized by the giant planet instability and trigger the Moon-forming impact.  When considered together with the initially dynamically unexcited models of \citet{desouza21}, we assess this particular result to be model-independent.

From our small sample of 56 simulations, we cannot rule out the possibility of an early ($t\simeq$ 50-100 Myr) instability completely reshaping a primordial inner solar system consisting of Earth, Venus, Theia and a few Mars analogs in a manner consistent with our initial hypothesis.  Several of the giant impacts in our simulations provide excellent analogs of the preferred Moon-forming impact geometries in hydrodynamical studies of the event \citep{canup04,canup12,cuk12}.  While systems finishing with a single Mars analog, and those generating a collisional fragment that survives to become a Mercury analog are both rare, such outcomes do occur (although they are mutually exclusive results within our small sample).  Thus, our results indicate that unlikely, but still possible that an early instability triggered the Moon-forming impact, removed unnecessary additional Mars analogs, and triggered a fragmenting collision that formed Mercury.  As our initial conditions are taken directly from the outcomes of late stage accretion models of the classic terrestrial planet formation scenario \citep{chambers98} originally reported in \citet{clement18_frag}, our work builds on the results of \citet{desouza21} who utilized idealized initial planetary configurations with low initial eccentricities and inclinations \citep[however, such initial conditions are potentially consistent with alternative terrestrial planet formation models][]{johansen21,izidoro21_nat,broz21}.

In spite of all efforts made, it is challenging for our systems to retain a planet in the Mars region.  As various studies of the Nice Model's effects on the fully formed terrestrial planets \citep{bras09,agnorlin12,bras13,kaibcham16} have highlighted the fragility of Mars' orbit during the epoch of giant planet migration, we see this as the strongest, model-independent case against our scenario.  While Earth and Venus' dynamically cold orbits are not produced in our models, this might be resolved if the planets pre-instability orbits were less excited \citep[e.g.:][]{johansen21,izidoro21_tp,izidoro21_nat}.  Mercury analogs are rare in our models, but it is possible that earlier processes or alternative disk conditions were responsible for the planets' mass and orbit \citep{clement21_merc2,clement21_merc4,broz21}.  While certain disk and instability scenarios have been shown to yield improved outcomes for Earth and Venus' excitation \citep{nesvorny21_tp}, the inability of our systems to consistently retain mass in the Mars region strongly speaks against our proposed scenario.

 The obvious solution for these shortcomings is an instability occurring during an earlier state of terrestrial evolution as suggested in the early instability model of \citet{clement18} and \citet{nesvorny21_tp}.  In this regard, it is noteworthy that the particular manner in which nebular gas dissipates has been interpreted as cause to necessitate an instability occurring in conjunction with gas dispersal \citep{liu22}.  However, our conclusion should be considered in the appropriate context given the fact that our work only considers a narrow range of initial parameters.  Thus, we refrain from making a wholesale claim of the applicability of our results to different terrestrial planet formation scenarios \citep[e.g.:][]{walsh11,ray17sci,bromley17,broz21,johansen21,izidoro21_tp,izidoro21_nat} or different initial terrestrial disk structures \citep{hansen09,iz14,lykawka19,clement21_merc2}.  It is also worth noting that triggering the Moon-forming impact is not necessarily a problem that requires a resolution, and appropriate collisions occur regularly in all of the potentially viable terrestrial planet formation scenarios.  Moreover, it is important to recognize that the vast majority of terrestrial formation models struggle to replicate Earth and Venus' eccentricities and inclinations when the effects of the Nice Model instability are accounted for.  Thus, it is possible that our models are failing to capture some key physical process that was responsible for damping the planets' orbits (for instance, residual photo-evaporating gas).

\section*{Acknowledgments}

The authors acknowledge the insightful reports of two anonymous reviewers that greatly improved the quality of the manuscript.  R.D. acknowledges support from the NASA EW program grant 80NSSC21K0387.  A.I. acknowledges NASA grant 80NSSC18K0828 to Rajdeep Dasgupta, during preparation and submission of the work.  A.I. also acknowledges support from the Welch Foundation grant No. C-2035-20200401.  The work described in this paper was supported by Carnegie Science's Scientific Computing Committee for High-Performance Computing (hpc.carnegiescience.edu).

\bibliographystyle{apj}
\newcommand{\sci}{$Science$ }
\newcommand{\psj}{$PSJ$}
\bibliography{tp_form.bib}

\appendix
\section{Supplemental Figures}

\begin{figure}
	\centering
	\includegraphics[width=.5\textwidth]{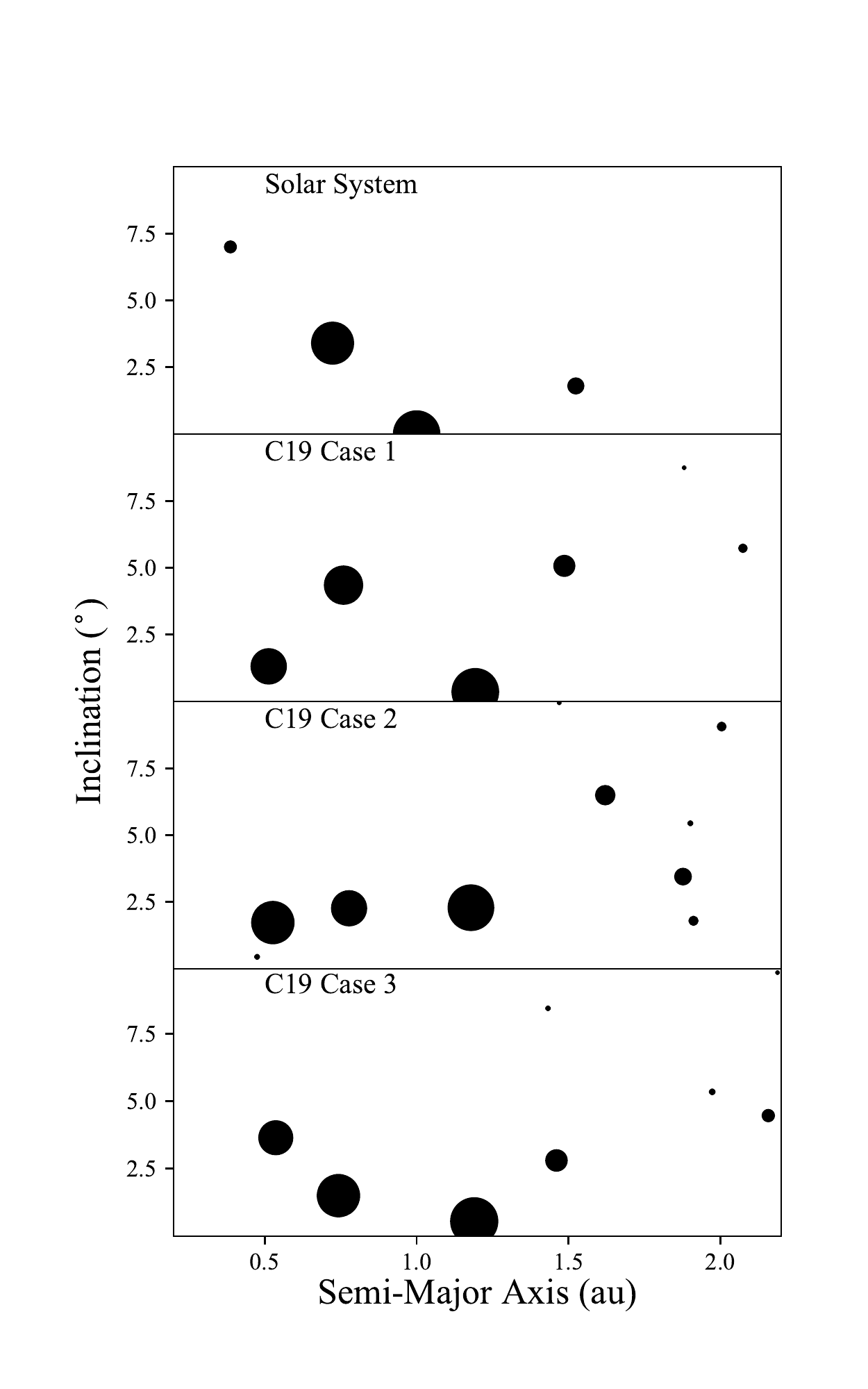}
	\caption{Same as figure \ref{fig:ics} except displaying inclination instead of eccentricity.}
	\label{fig:ics_inc}
\end{figure}

\begin{figure}
	\centering
	\includegraphics[width=.5\textwidth]{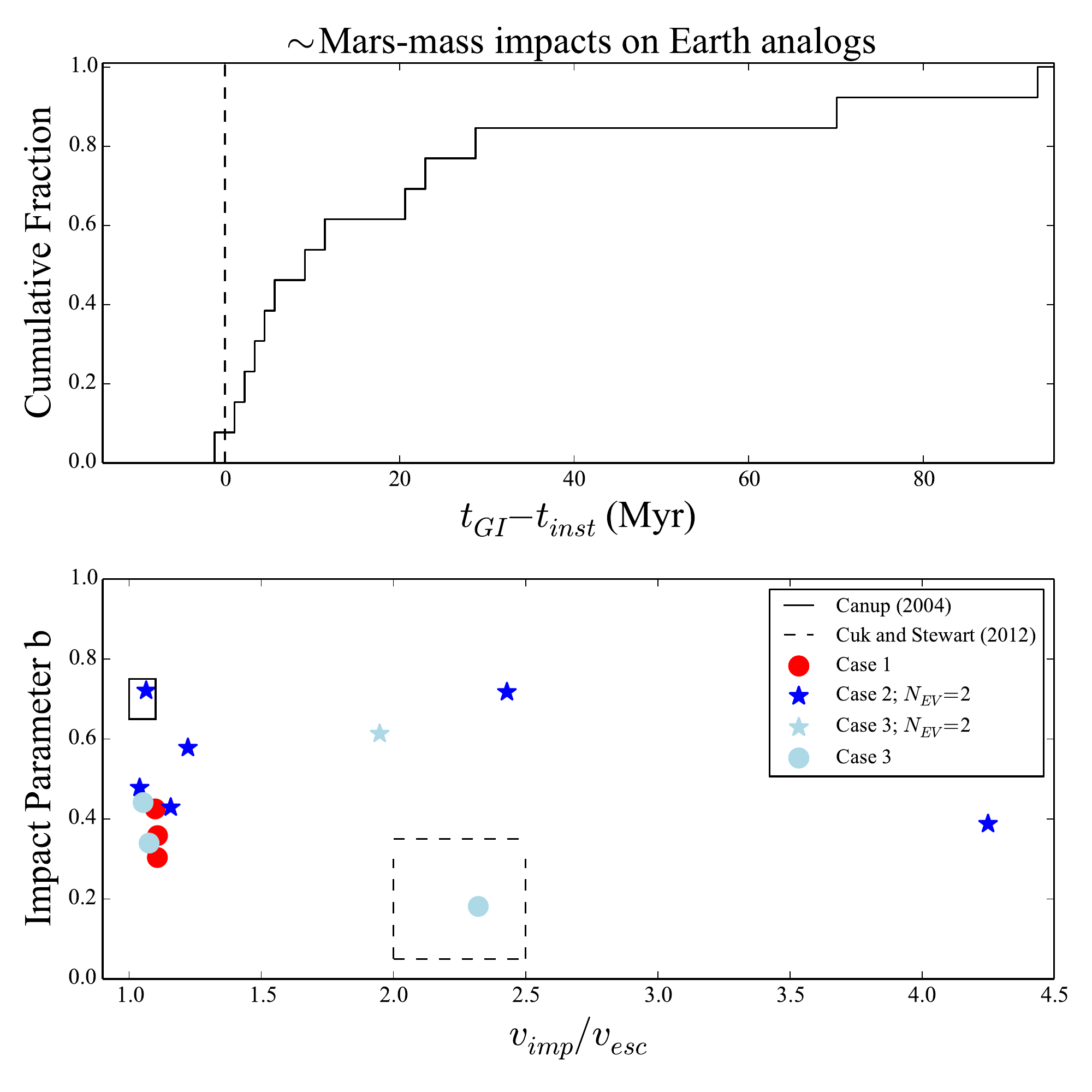}
	\caption{Same as figure \ref{fig:moon}, except here we plot impacts between Earth analogs (the outermost of three initial massive proto-planets in figure \ref{fig:ics}) and Mars analogs (0.025 $<m<$ 0.3 $M_{\oplus}$) in systems where the inner two proto-planets (Venus and Theia) merge or survive on stable orbits without experiencing a giant impact.  Systems annotated with stars finish with exactly two Earth/Venus analogs.  The solid box displays the preferred parameters from the analysis of the canonical impact scenario in \citet{canup04}.  Similarly, the dashed line box roughly marks the region of favored impacts from the study of \citet{cuk12} that favored a higher-velocity impact followed by angular momentum reduction of the Earth-Moon system through the evection resonance with the Sun.}
	\label{fig:mars} 
\end{figure}

\end{document}